\documentclass[10pt.draft]{article}
\usepackage{amssymb}
\usepackage{graphicx}
\usepackage{amsmath}
\usepackage{acronym}

\input{tcilatex}

\begin{document}

\bigskip

\ {\huge New models for Veneziano }

\ {\huge amplitudes}\ {\Huge :}{\huge \ \ Combinatorial, }

\ {\huge symplectic and supersymmetric}

{\huge \ \ \ \ \ \ \ \ \ \ \ \ \ \ \ \ aspects}

$\ \ \ \ \ \ \ \ \ \ \ \ \ \ \ \ \ \ \ \ \ \ \ \ \ \ \ \ \ \ \ \ \ $

\ \ \ \ \ \ \ \ \ \ \ \ \ \ \ \ \ \ \ \ \ \ \ \ \ \ \ A.L. Kholodenko%
\footnote{%
E-mail address: string@clemson.edu}

\textit{375 H.L.Hunter Laboratories, Clemson University, Clemson, }

\textit{SC} 29634-0973, USA

\bigskip

The bosonic string theory evolved as an attempt to find a physical/quantum
mechanical model \ capable of reproducing Euler's beta function (Veneziano
amplitude) and its multidimensional analogue. The multidimensional analogue
of beta function was studied mathematically for some time from different
angles by mathematicians such as Selberg,Weil and Deligne among many others.
The results of their studies apparently were not taken into account in
physics literature on string theory. In several recent publications attempts
were made to restore the missing links. As discussed in these publications,
the existing mathematical interpretation of the multidimensional analogue of
Euler's beta function as one of the periods associated with the
corresponding differential form \textquotedblright living\textquotedblright\
on the Fermat-type (hyper) surface, happens to be crucial for restoration of
the quantum/statistical mechanical models reproducing such generalized beta
function. There is a number of \ nontraditional models -all interrelated-
capable of reproducing the Veneziano amplitudes. In this work we would like
to discuss two of such new models: symplectic and supersymmetric. The
symplectic model is based on observation that the Veneziano amplitude is
just the Laplace transform of the generating function for the Ehrhart
polynomial. Such a polynomial counts the number of lattice points inside the
rational polytope \ (i.e. polytope whose vertices are located at the nodes
of a regular lattice) and at its boundaries. In the present case the
polytope is a regular simplex. It is a deformation retract for the
Fermat-type (hyper)surface (perhaps inflated, as explained in the text).
Using known connections between polytopes and dynamical systems the quantum
mechanical system associated with such a dynamical system is found. The
ground state of this system is degenerate with degeneracy factor given by
the Ehrhart polynomial. \ Using some ideas by Atiyah, Bott and Witten we
argue that the supersymmetric model related to the symplectic can be
recovered. While recovering this model, we demonstrate that the ground state
of such a model is degenerate with the same degeneracy factor as for earlier
obtained symplectic model. Since the wave functions of this model are in one
to one correspondence with the Veneziano amplitudes, this exactly solvable
supersymmetric (and, hence, also symplectic) model is sufficient for
recovery of the partition function reproducing the Veneziano amplitudes thus
providing the exact solution of the Veneziano model.

\textit{Kewords}: Veneziano and Veneziano-like amplitudes; Ehrhart
polynomial for rational polytopes; Duistermaat-Heckman formula;
Khovanskii-Pukhlikov correspondence; Lefschetz isomorphism theorem.

\pagebreak

\ \ \ 

\section{\protect\bigskip Introduction}

\bigskip In 1968 Veneziano [1] postulated the 4-particle scattering
amplitude $A(s,t,u)$ given (up to a common constant factor) by 
\begin{equation}
A(s,t,u)=V(s,t)+V(s,u)+V(t,u),  \tag{1}
\end{equation}%
where 
\begin{equation}
V(s,t)=\int\limits_{0}^{1}x^{-\alpha (s)-1}(1-x)^{-\alpha (t)-1}dx\equiv
B(-\alpha (s),-\alpha (t))  \tag{2}
\end{equation}%
is the Euler beta function and $\alpha (x)$ is the Regge trajectory usually
written as $\alpha (x)=\alpha (0)+\alpha ^{\prime }x$ with $\alpha (0)$ and $%
\alpha ^{\prime }$ being the Regge slope and intercept, respectively. In
case of space-time metric with signature $\{-,+,+,+\}$ the Mandelstam
variables $s$, $t$ and $u$ entering the Regge trajectory are defined by [2] 
\begin{equation}
s=-(p_{1}+p_{2})^{2};\text{ }t=-(p_{2}+p_{3})^{2};\text{ }%
u=-(p_{3}+p_{1})^{2}.  \tag{3}
\end{equation}%
The 4-momenta $p_{i}$\ are constrained by the energy-momentum conservation
law leading to relation between the Mandelstam variables: 
\begin{equation}
s+t+u=\sum\limits_{i=1}^{4}m_{i}^{2}.  \tag{4}
\end{equation}%
Veneziano [1] noticed\footnote{%
To get our Eq.(5) from Eq.7 of Veneziano paper, it is sufficient to notice
that his $1-\alpha (s)$ corresponds to ours -$\alpha (s).$} that to fit
experimental data the Regge trajectories should obey the constraint 
\begin{equation}
\alpha (s)+\alpha (t)+\alpha (u)=-1  \tag{5}
\end{equation}%
consistent with Eq.(4) in view of the definition of $\alpha (s).$ The
Veneziano condition, Eq.(5),\ can be rewritten in a more general form.
Indeed, let $m,n,l$ be some integers such that $\alpha (s)m+\alpha
(t)n+\alpha (u)l=0$. Then by adding this equation to Eq.(5) we obtain, $%
\alpha (s)\tilde{m}+\alpha (t)\tilde{n}+\alpha (u)\tilde{l}=-1,$ or more
generally, $\alpha (s)\tilde{m}+\alpha (t)\tilde{n}+\alpha (u)\tilde{l}+%
\tilde{k}\cdot 1=0.$ Both equations have been studied extensively in the
book by Stanley [3] from the point of view of commutative algebra,
polytopes, toric varieties, invariants of finite groups, etc. This
observation plays the major role in developments we would like to present in
this work and elsewhere.

In 1967-a year before Veneziano's paper was published- the paper [4] by
Chowla and Selberg appeared relating Euler's beta function to the periods of
elliptic integrals. The result by Chowla and Selberg was generalized by
Andre Weil whose two influential papers [5,6] brought into the picture the
periods of Jacobians of the Abelian varieties, Hodge rings, etc. Being
motivated by these papers, Benedict Gross wrote a paper [7] in which the
beta function appears as period associated with the differential form
\textquotedblright living\textquotedblright\ on the Jacobian of the Fermat
curve. His results as well as those by Rohrlich (placed in the appendix to
Gross paper) have been subsequently documented in the book by Lang [8].
Although in the paper by Gross the multidimensional extension of beta
function is considered only briefly, e.g.[7,p.207], the computational
details were not provided hovewer. These computational details can be found
in our recently published papers, Refs.[9,10]. To obtain the
multidimensional extension of beta function as period integral, following
logic of papers by Gross and Deligne [11], one needs to replace the Fermat
curve by the Fermat hypersufrace, to embed it into the complex projective
space, and to treat it as K\"{a}hler manifold. The differential forms living
on such manifold are associated with periods of the Fermat hypersurface.
Physical considerations require this K\"{a}hler manifold to be of the Hodge
type. In his lecture notes [11] Deligne noticed that the Hodge theory needs
some essential changes (e.g. mixed Hodge structures, etc.) if the Hodge-K%
\"{a}hler manifolds possess singularities. Such modifications may be needed
upon development of our formalism. A monograph by Carlson et al, Ref.[12],
contains an up to date exaustive information regarding such modifications,
etc. Fortunately, to obtain the multiparticle Veneziano amplitudes these
complications are not nesessary. In Ref. [10] we demonstrated that the
period integrals living on the Fermat hypersurfaces, when properly
interpreted, provide the tachyon-free (Veneziano-like) multiparticle
amplitudes whose paricle spectrum reproduces those known for both the open
and closed bosonic strings. Naturally, the question arises: If this is so,
then what kind of a model is capable of reproducing such amplitudes ?\ In
this paper we would like to discuss some combinatorial properties of the
Veneziano amplitudes \ (easily extendable to the Veneziano-like) sufficient
for reproducing \ at least two of such models: symplectic and
supersymmetric. Mathematically, the results presented below are in accord
with those by Vergne [13] whose work does not contain practical applications.

This work is organized as follows. In Section 2 we explain the combinatorial
nature of the Veneziano amplitudes by connecting them with the generating
function for the Ehrhart polynomial. Such a polynomial counts the number of
points inside the rational polytope (i.e. \ polytope whose vertices are
located at the nodes of the regular $k-$dimensional lattice) and at its
boundaries (faces). In the present case the polytope is a regular simplex
which is a deformation retract for the Fermat-type (hyper) surface living in
the complex projective space [9,10]. The connections between the polytopes
and dynamical systems are well-known [13,14]. Development of these
connections is presented in Sections 2-4 where we find the corresponding
quantum mechanical system whose ground state is degenerate with degeneracy
factor being identified with the Ehrhart polynomial. The obtained result is
in accord with that by Vergne [13]. In addition, in Section 5 the generating
function for the Ehrhart polynomial is reinterpreted in terms of the Poincar$%
e^{\prime }$ polynomial. Such a polynomial is used, for instance, in the
theory of invariants of finite (pseudo)reflection groups [3,15]. Obtained
indentification reveals the topological and group-theoretic nature of the
Veneziano amplitudes. To strengthen this point of view, we use some results
by Atiyah and Bott [16] inspired by still earlier work by Witten [17] on
supersymmetric quantum mechanics. They allow us to think about the Veneziano
amplitudes using therminology of intersection theory [18]. This is
consistent with earlier mentioned interpretation of these amplitudes in
terms of periods of the Fermat (hyper)surface [19]. It also makes
computation of these amplitudes analogous to those for the Witten
-Kontsevich model [20, 21], whose refinements can be found in our earlier
work, Ref.[22]. For the sake of space, in this work we do not develop these
connections with the Witten- Kontsevich model any further. Instead, we
discuss the supersymmetric model associated with symplectic model described
earlier and treat it with help of the Lefshetz isomorphism theorem. This
allows us to look at the problem of computation of the spectrum for such a
model from the point of view of the theory of representations of the complex
semisimple Lie algebras. Using some results by Serre [23] and Ginzburg [24]
we demonstrate that the ground state for such finite dimensional
supersymmetric quantum mechanical model is degenerate with degeneracy factor
coinciding with the Erhardt polynomial. This result is consistent with that
obtained in Section 4 by different methods.

\section{ The Veneziano amplitude and the Ehrhart polynomial}

\bigskip

In view of Eq.(2), consider an identity taken from [25],

\begin{align}
\frac{1}{(1-tz_{0})\cdot \cdot \cdot (1-tz_{k})}& =(1+tz_{0}+\left(
tz_{0}\right) ^{2}+...)\cdot \cdot \cdot (1+tz_{n}+\left( tz_{n}\right)
^{2}+...)  \notag \\
& =\sum\limits_{n=0}^{\infty
}(\sum\limits_{k_{0}+...+k_{k}=n}z_{0}^{k_{0}}\cdot \cdot \cdot
z_{k}^{k_{k}})t^{n}.  \tag{6}
\end{align}%
When $z_{0}=...=z_{k}=1,$the inner sum in the last expression provides the
total number of monomials of the type $z_{0}^{k_{0}}\cdot \cdot \cdot
z_{k}^{k_{k}}$ with $k_{0}+...+k_{k}=n$. The total number of such monomials
is given by the binomial coefficient\footnote{%
The reason for displaying 3 different forms of the same combinatorial factor
will be explained shortly below.} 
\begin{equation}
p(k,n)=\frac{\left( k+n\right) !}{k!n!}=\frac{(n+1)(n+2)\cdot \cdot \cdot
(n+k)}{k!}=\frac{(k+1)(k+2)\cdot \cdot \cdot (k+n)}{n!}.  \tag{7}
\end{equation}%
For this special case (6) is converted to a useful expansion, 
\begin{equation}
P(k,t)\equiv \frac{1}{\left( 1-t\right) ^{k+1}}=\sum\limits_{n=0}^{\infty
}p(k,n)t^{n}.  \tag{8}
\end{equation}%
In view of the integral representation of the beta function given by (2), \
we replace $k+1$ by $\alpha (s)+1$ in (8) and use it in the beta function
representation of $V(s,t)$. Straightforward calculation produces the
following known result [2]: 
\begin{equation}
V(s,t)=-\sum\limits_{n=0}^{\infty }p(\alpha (s),n)\frac{1}{\alpha (t)-n}. 
\tag{9}
\end{equation}%
The r.h.s. of (9) is effectively the Laplace transform of the generating
function (8) which can be interpreted as a partition function in the sence
of statistical mechanics. The purpose of this Letter is to demonstrate that
such an interpretation is not merely a conjecture and, in view of this, to
find the statisical mechanical/quantum model whose partition function is
given by Eq.(8). \ Our arguments are not restricted to the 4-particle
amplitude. Indeed, as we argued earlier [10], the multidimensional extension
of Euler's beta function producing murtiparticle Veneziano amplitudes (upon
symmetrization analogous to the 4-particle case) is given by the following
integral attributed to Dirichlet 
\begin{equation}
\mathcal{D}(x_{1},...,x_{k})=\int \int\limits_{\substack{ u_{1}\geq
0,...,u_{k}\geq 0  \\ u_{1}\text{ }+\cdot \cdot \cdot +u_{k}\leq 1}}\unit{u}%
_{1}^{x_{1}-1}\unit{u}_{2}^{x_{2}-1}...\unit{u}%
_{k}^{x_{k}-1}(1-u_{1}-...-u_{k})^{x_{k+1}-1}du_{1}...du_{k}.  \tag{10}
\end{equation}%
In this integral let $t=u_{1}+...+u_{k}$. This allows us to use already
familiar expansion (8). In addition, the following identity 
\begin{equation}
t^{n}=(u_{1}+...+u_{k})^{n}=\sum\limits_{n=(n_{1},...,n_{k})}\frac{n!}{%
n_{1}!n_{2}!...n_{k}!}u_{1}^{n_{1}}\cdot \cdot \cdot u_{k}^{n_{k}}  \tag{11}
\end{equation}%
with restriction $n=n_{1}+...+n_{k}$ is of importance as well. This type of
identity was used earlier in our work on Kontsevich-Witten model [22].
Moreover, from the same paper it follows that the above result can be
presented as well in the alternative useful form: 
\begin{equation}
(u_{1}+...+u_{k})^{n}=\sum\limits_{\lambda \vdash k}f^{\lambda }S_{\lambda
}(u_{1},...,u_{k}),  \tag{12}
\end{equation}%
where the Schur polynomial $S_{\lambda }$ is defined by 
\begin{equation}
S_{\lambda }(u_{1},...,u_{k})=\sum\limits_{n=(n_{1},...,n_{k})}K_{\lambda
,n}u_{1}^{n_{1}}\cdot \cdot \cdot u_{k}^{n_{k}}  \tag{13}
\end{equation}%
with coefficients $K_{\lambda ,n}$ known as Kostka numbers, $\ f^{\lambda }$
being the number of standard Young tableaux of shape $\lambda $ and the
notation $\lambda \vdash k$ meaning that $\lambda $ is partition of $k$.
Through such a connection with Schur polynomials one can develop connections
with the Kadomtsev-Petviashvili (KP) hierarchy of nonlinear exactly
integrable systems on one hand\ and with the theory of Schubert varieties on
another. \ Although details \ can be found in our earlier work [22], in this
Letter we shall discuss these issues a bit further in Section\textbf{\ }5%
\textbf{. }Use of (11) in (10) produces, after performing the multiple
Laplace transform, the following part of the multiparticle Veneziano
amplitude 
\begin{equation}
A(1,...k)=\frac{\Gamma _{n_{1}...n_{k}}(\alpha (s_{k+1}))}{(\alpha
(s_{1})-n_{1})\cdot \cdot \cdot (\alpha (s_{k})-n_{k})}.  \tag{14}
\end{equation}%
Even though the residue $\Gamma _{n_{1}...n_{k}}(\alpha (s_{k+1}))$ contains
all the combinatorial factors, the obtained result should still be
symmetrized (in accord with the 4-particle case considered by Veneziano) in
order to obtain the full murtiparticle Veneziano amplitude. Since in the
above general multiparticle case the same expansion (8) was used,\ for the
sake of space it is sufficient to focus on the 4-particle amplitude only.
This task is reduced to further study of the expansion (8). Such an
expansion can be looked upon from several different angles. For instance, we
have mentioned already that it can be interpreted as a partition function.
In addition, it is the generating function for the Ehrhart polynomial. The
combinatorial factor $p(k,n)$ defined in (7) is the simplest example of the
Ehrhart polynomial. Evidently, it can be written formally as 
\begin{equation}
p(k,n)=a_{n}k^{n}+a_{n-1}k^{n-1}+\cdot \cdot \cdot +a_{0}.  \tag{15}
\end{equation}%
Let $\mathcal{P}$ be \textit{any} convex rational polytope that is the
polytope whose vertices are located at the nodes of some $n-$dimensional 
\textbf{Z}$^{n}$ lattice. Then, the Ehrhart polynomial \ for the inflated
polytope $\mathcal{P}$ (with coefficient of inflation\textit{\ }$\mathit{%
k=1,2,..}$\textit{.}) can be written as 
\begin{equation}
\left\vert k\mathcal{P}\cap \mathbf{Z}^{n}\right\vert =\mathfrak{P}%
(k,n)=a_{n}(\mathcal{P})k^{n}+a_{n-1}(\mathcal{P})k^{n-1}+\cdot \cdot \cdot
+a_{0}(\mathcal{P})  \tag{16}
\end{equation}%
with coefficients $a_{0},...,a_{n}$ being specific for a given type of
polytope $\mathcal{P}$. In the case of Veneziano amplitude the polynomial $%
p(k,n)$ counts number of points inside the $n-$dimensional inflated simplex
(with inflation coefficient $k=1,2,...$). Irrespective to the polytope type,
it is known [26] that $a_{0}=1$ and $a_{n}=Vol\mathcal{P},$where $Vol%
\mathcal{P}$ is the $\mathit{Euclidean}$ volume of the polytope. These facts
can be easily checked for $p(k,n).$ To calculate the remaining coefficients
of such polynomial explicitly for arbitrary convex rational polytope $%
\mathcal{P}$ is a difficult task in general. Such a task was accomplished
only recently in [27]. The authors of [27] recognized that in order to
obtain the remaining coefficients, it is useful to calculate the generating
function for the Ehrhart polynomial. Long before the results of [27] were
published, it was known [3\textbf{,}15], that the generating function for
the Ehrhart polynomial of $\mathcal{P}$ can be written in the following
universal form 
\begin{equation}
\mathcal{F}(\mathcal{P},x)=\sum\limits_{k=0}^{\infty }\mathfrak{P}(k,n)x^{k}=%
\frac{h_{0}(P)+h_{1}(P)x+\cdot \cdot \cdot +h_{n}(P)x^{n}}{(1-x)^{n+1}} 
\tag{17}
\end{equation}%
The above general result \ might be of some use in case of possible
generalizations of the Veneziano amplitudes and the associated with them
partition functions. Its mathematical meaning is discussed further in
Section 5.

The fact that the combinatorial factor $p(k,n)$ in (7) can be formally
written in several equivalent ways has some physical significance. For
instance, in particle physics literature, e.g. see [2], the second option is
commonly used. Let us recall how this happens. One is looking for an
expansion of the factor $(1-x)^{-\alpha (t)-1}$ under the integral of beta
function, e.g. see Eq.(2). \ Looking at Eq.(17) one realizes that the
Mandelstam variable $\alpha (t)$ plays the role of dimensionality of \textbf{%
Z-} lattice. Hence, in view of (6), we have to identify it with $n$ in the
second option provided by (7). This is not the way such an identification is
done in physics literature where, in fact, the third option\ in (7) is
commonly used with $n=\alpha (t)$ \ being effectively the inflation factor
while $k$ \ is effectively the dimensionality of the lattice.\footnote{%
We have to warn our readers that nowhere in physics literature such
combinatorial terminology is used.} A quick look at Eq.s(8) and (17) shows
that under such circumstances the generating function for the Ehrhart
polynomial and that for the Veneziano amplitude are formally not the same.
In the first case one is dealing with lattices of \textit{fixed}
dimensionality and considering summation over various inflation factors at
the same time. In the second (physical) case, one is dealing with the 
\textit{fixed inflation factor} $n=\alpha (t)$ while summing over lattices
of different dimensionalities. Such arguments are superficial however in
view of \ Eq.s(6) and (17) above.Using these equations it is clear that
mathematically correct agreement between Eq.s(8) and (17) can be reached if
one is using $\mathfrak{P}(k,n)=p(k,n)$ with the second option taken from
(7). By doing so no changes in the pole locations for the Veneziano
amplitude occur. Moreover, for a given pole the second and the third option
in (7) produce exactly the same contributions into the residue thus making
them physically indistinguishable. The interpretation of the Veneziano
amplitude as the Laplace transform of the Ehrhart polynomial generating
function provides a very compelling reason for development of the
alternative string-theoretic formalism. In addition, it allows us to think
about possible generalizations of the Veneziano amplitude using generating
functions for the Ehrhart polynomials for polytopes other than the $n-$%
dimensional inflated simplex used for the Veneziano amplitudes. As it is
demonstrated by Stanley [3,15], Eq.(17) has a group invariant meaning as the
Poincare$^{\prime }$ polynomial for the so called Stanley-Reisner polynomial
ring.\footnote{%
In Section 5 we provide some additional details on this topic.} From the
same reference one can also find connections of these results with toric
varieties. \ In view of \ Ref. [14], this observation is sufficient for
restoration of physical models reproducing the Veneziano and Veneziano-like
amplitudes. Thus, in the rest of this paper we shall discuss some approaches
to the design of these models. Space limitations forbid us from discussing
other directions. They will be discussed in the subsequent publications.

\section{\protect\bigskip\ Motivating examples}

To facilitate our readers understanding, we would like to illustrate general
principles using simple examples. We begin by considering a finite geometric
progression of the type 
\begin{align}
\mathcal{F(}c,m)& =\sum\limits_{l=-m}^{m}\exp \{cl\}=\exp
\{-cm\}\sum\limits_{l=0}^{\infty }\exp \{cl\}+\exp
\{cm\}\sum\limits_{l=-\infty }^{0}\exp \{cl\}  \notag \\
& =\exp \{-cm\}\frac{1}{1-\exp \{c\}}+\exp \{cm\}\frac{1}{1-\exp \{-c\}} 
\notag \\
& =\exp \{-cm\}\left[ \frac{\exp \{c(2m+1)\}-1}{\exp \{c\}-1}\right] . 
\tag{18}
\end{align}%
The reason for displaying the intermediate steps will be explained shortly
below. First, however, we would like to consider the limit : $c\rightarrow
0^{+}$ of $\mathcal{F(}c,m)$. It is given by $\mathcal{F(}0,m)=2m+1$. The
number $2m+1$ equals to the number of integer points in the segment $[-m,m]$ 
\textit{including} \textit{boundary} points. It is convenient to rewrite the
above result in terms of $x=\exp \{c\}$ so that we shall write formally $%
\mathcal{F(}x,m)$ instead of $\mathcal{F(}c,m)$ from now on. Using such
notation, consider a related function 
\begin{equation}
\mathcal{\bar{F}(}x,m)=(-1)\mathcal{F(}\frac{1}{x},-m).  \tag{19}
\end{equation}%
This type of relation (the \textit{Ehrhart-Macdonald reciprocity law}) is
characteristic for the Ehrhart polynomial for rational polytopes discussed
earlier. In the present case we obtain, 
\begin{equation}
\mathcal{\bar{F}(}x,m)=(-1)\frac{x^{-(-2m+1)}-1}{x^{-1}-1}x^{m}.  \tag{20}
\end{equation}%
In the limit $x\rightarrow 1+0^{+}$ we obtain : $\mathcal{\bar{F}(}%
1,m)=2m-1. $ The number $2m-1$ is equal to the number of integer points
strictly \textit{inside} the segment $[-m,m].$ \ Both $\mathcal{F(}0,m)$ and 
$\mathcal{\bar{F}(}1,m)$ \ provide the simplest possible examples of the
Ehrhart polynomials if we identify $m$ with the inflation factor $k$.

These, seemingly trivial, results can be broadly generalized. First, we
replace $x$ by $\mathbf{x=}x_{1}\cdot\cdot\cdot x_{d}$, next we replace the
summation sign in the left hand side of Eq.(18) by the multiple summation,
etc. Thus obtained function $\mathcal{F(}\mathbf{x},m)$ in the limit $%
x_{i}\rightarrow1+0^{+},$ $i=1-d,$ produces the anticipated result $:%
\mathcal{F(}\mathbf{1},m)=(2m+1)^{d}$ . It describes the number of points
inside and at the faces of $\ $a $d-$ dimensional cube in the Euclidean
space $\mathbf{R}^{d}$. Accordingly, for the number of points strictly
inside the cube we obtain : $\mathcal{\bar{F}(}\mathbf{1},m)=(2m-1)^{d}.$

To move further, we would like to remind to our readers that \
mathematically a subset of \ $\mathbf{R}^{n}$ is considered to be a \textit{%
polytope (or polyhedron)} $\mathcal{P}$ if there is a $r\times d$ matrix $%
\mathbf{M}$\textbf{\ (}with\textbf{\ }$r\leq d)$ and a vector $\mathbf{b}\in 
\mathbf{R}^{d}$ such that $\mathcal{P}=\{\mathbf{x}\in \mathbf{R}^{d}\mid 
\mathbf{Mx}\leq \mathbf{b}\}.$ Provided that the Euclidean $d$-dimensional
scalar product is given by $<\mathbf{x}\cdot \mathbf{y}>=\sum%
\limits_{i=1}^{d}x_{i}y_{i}$ , a \textit{rational (or integral) polytope (or
polyhedron)} $\mathcal{P}$ is defined by the set 
\begin{equation}
\mathcal{P}=\{\mathbf{x}\in \mathbf{R}^{d}\mid <\mathbf{a}_{i}\cdot \mathbf{x%
}>\leq \beta _{i}\text{ , }i=1,...,r\}  \tag{21}
\end{equation}%
where $\mathbf{a}_{i}\in \mathbf{Z}^{n}$ and $\beta _{i}\in \mathbf{Z}$ for $%
i=1,...,r.$ Let \textit{Vert}$\mathcal{P}$ denote the vertex set of the
rational polytope, in the case considered thus far, the $d-$dimensional
cube. Let $\{u_{1}^{v},...,u_{d}^{v}\}$ denote the orthogonal basis (not
necessarily of unit length) made of the highest weight vectors of the
Weyl-Coxeter reflection group $B_{d}$ appropriate for the cubic symmetry
[28].These vectors are oriented along the positive semi axes with respect to
the center of symmetry of the cube. When parallel translated to the edges
ending at particular hypercube vertex \textbf{v}, they can point either in
or out of this vertex. Then, the $d$-dimensional version of Eq.(18) can be
rewritten in notations just introduced as follows 
\begin{equation}
\sum\limits_{\mathbf{x}\in \mathcal{P\cap }\mathbf{Z}^{d}}\exp \{<\mathbf{c}%
\cdot \mathbf{x}>\}=\sum\limits_{\mathbf{v}\in Vert\mathcal{P}}\exp \{<%
\mathbf{c}\cdot \mathbf{v}>\}\left[ \prod\limits_{i=1}^{d}(1-\exp
\{-c_{i}u_{i}^{v}\})\right] ^{-1}.  \tag{22}
\end{equation}%
The correctness of this equation can be readily checked by considering
special cases of a segment, square, cube, etc. The result, Eq.(22), obtained
for the polytope of cubic symmetry can be extended to the arbitrary convex
centrally symmetric polytope. Details can be found in Ref.[29]. Moreover,
the requirement of central symmetry can be relaxed to the requirement of the
convexity of \ $\mathcal{P}$ only. In such general form the relation given
by Eq.(22) was obtained by Brion [30]. It is of central importance for the
purposes of this work: the limiting procedure $c\rightarrow 0^{+}$ produces
the number of points inside (and at the boundaries) of the polyhedron $%
\mathcal{P}$ in the l.h.s. of Eq.(22) and, if the polyhedron is rational and
inflated, this procedure produces the Ehrhart polynomial. Actual
computations are done with help of the r.h.s. of Eq.(22) as will be
demonstrated below.

\section{\protect\bigskip\ The Duistermaat-Heckman formula and the
Khovanskii-Pukhlikov correspondence}

Since the description of the Duistermaat-Heckman \ (D-H) formula can be
found in many places, we would like to be brief \ in discussing it now in
connection with earlier obtained results. Let $M\equiv M^{2n}$ be a compact
symplectic manifold equipped with the moment map $\Phi :M\rightarrow \mathbf{%
R}$ and the (Liouville) volume form $dV=\left( \frac{1}{2\pi }\right) ^{n}%
\frac{1}{n!}$ $\Omega ^{n}.$\ According to the Darboux theorem, locally $%
\Omega =\sum\nolimits_{l=1}^{n}$\ \ $dq_{l}$\ $\wedge dp_{l}$\ . We expect
that such a manifold has isolated fixed points $p$ belonging to the fixed
point set $\mathcal{V}$ associated with the isotropy subgroup of the group $%
G $ acting on $M$. Then, in its most general form, the D-H formula can be
written as [13,14,31] 
\begin{equation}
\int\limits_{M}dVe^{\Phi }=\sum\limits_{p\in \mathcal{V}}\frac{e^{\Phi (p)}}{%
\prod\nolimits_{j}a_{j,p}}  \tag{23}
\end{equation}
where $a_{1,p},...,a_{n,p}$ are the weights of the linearized action of $G$
on $T_{p}M$ . Using Morse theory, Atiyah [32] and others, e.g. see Ref.[14]
for additional references, have demonstrated that it is sufficient to keep
terms up to quadratic in the expansion of $\Phi $ around given $p$. In such
a case the moment map can be associated with the Hamiltonian for the finite
set of harmonic oscillators. In the properly chosen system of units the
coefficients $a_{1,p},...,a_{n,p}$ are just ''masses''\ $m_{i}$ \ of the
individual oscillators. Unlike truly physical masses, some of $m_{i}^{\prime
}s$ can be negative.

Based on the information just provided, we would like be more specific now.
To this purpose, following Vergne [33] and Brion [30], we would like to
consider the D-H integral of the form 
\begin{equation}
I(k\text{ };\text{ }y_{1},y_{2})=\int\nolimits_{k\Delta }dx_{1}dx_{2}\exp
\{-(y_{1}x_{1}+y_{2}x_{2})\},  \tag{24}
\end{equation}%
where $k\Delta $ is the standard dilated simplex with \ dilation coefficient 
$k\footnote{%
Our choice of the simplex as the domain of integration is caused by our
earlier made observation [10] that the deformation retract of the Fermat
(hyper)surface (on which the Veneziano amplitude lives ) is just the
standard simplex. Since such Fermat surface is a complex K\"{a}hler-Hodge
type manifold and since all K\"{a}hler manifolds are symplectic [14,34], our
choice makes sense.}$. Calculation of \ this integral can be done exactly
with the result: 
\begin{equation}
I(k\text{ };\text{ }y_{1},y_{2})=\frac{1}{y_{1}y_{2}}+\frac{e^{-ky_{1}}}{%
y_{1}(y_{1}-y_{2})}+\frac{e^{-ky_{2}}}{y_{2}(y_{2}-y_{1})}  \tag{25}
\end{equation}%
consistent with Eq.(23). In the limit: $y_{1},y_{2}\rightarrow 0$ some
calculation produces the anticipated result $:Volk\Delta =k^{2}/2!$ \ for
the \textit{Euclidean} volume of the dilated simplex. Next, to make a
connection with the previous section, in particular, with Eq.(22), consider
the following sum 
\begin{align}
S(k\text{ ; }y_{1}\text{,}y_{2})& =\sum\limits_{\left( l_{1},l_{2}\right)
\in k\Delta }\exp \{-(y_{1}l_{1}+y_{2}l_{2})\}  \notag \\
& =\frac{1}{1-e^{-y_{1}}}\frac{1}{1-e^{-y_{2}}}+\frac{1}{1-e^{y_{1}}}\frac{%
e^{-ky_{1}}}{1-e^{y_{1}-y_{2}}}+\frac{1}{1-e^{y_{2}}}\frac{e^{-ky_{2}}}{%
1-e^{y_{2}-y_{1}}}  \tag{26}
\end{align}%
related to the D-H integral, Eq.(24). Its calculation will be explained
momentarily. In spite of the connection with the D-H integral, the limiting
procedure: $y_{1},y_{2}\rightarrow 0$ in the last case is much harder to
perform. It is facilitated by use of the following expansion 
\begin{equation}
\frac{1}{1-e^{-s}}=\frac{1}{s}+\frac{1}{2}+\frac{s}{12}+O(s^{2}).  \tag{27}
\end{equation}%
Rather lenghty calculations produce the anticipated result : $S(k$ ; $0$,$%
0)=k^{2}/2!+3k/2+1\equiv \left\vert k\Delta \cap Z^{2}\right\vert \equiv 
\mathfrak{P}(k,2)$ for the Ehrhart polynomial. Since generalization of the
obtained results to simplicies of higher dimensions is straightforward, the
relevance of these results to the Veneziano amplitude should be evident. To
make it more explicit we have to make several steps still. First, we would
like to explain how the result (26) was obtained. By doing so we shall gain
some additional physical information. Second, we would like to explain in
some detail the connection between the integral (25), and the sum (26). Such
a connection is made with help of the Khovanskii-Pukhlikov correspondence.

We begin with calculations of the sum, Eq.(26). To do this we need a
definition of the convex rational polyhedral cone $\sigma .$ It is given by 
\begin{equation}
\sigma =\mathbf{Z}_{\geq 0}a_{1\text{\ }}+\cdot \cdot \cdot +\mathbf{Z}%
_{\geq 0}a_{d\text{\ }},  \tag{28}
\end{equation}%
where the set $a_{1},...,a_{d}$ \ forms a basis (not nesessarily orthogonal)
of the $d$-dimensional vector space $V,$ while $\mathbf{Z}_{\geq 0}$ are non
negative integers. It is known that all combinatorial information about the
polytope $\mathcal{P}$ is encoded in the $\mathit{complete}$ fan made of
cones whose apexes all having the same origin in common. Details can be
found in literature [14,18]. At the same time, the vertices of $\mathcal{P}$
are also the apexes of the respective cones. Following Brion[30] this fact
allows us to write the l.h.s. of Eq.(22) as 
\begin{equation}
f(\mathcal{P},x)=\sum\limits_{\mathbf{m}\in \mathcal{P}\cap \mathbf{Z}^{d}}%
\mathbf{x}^{\mathbf{m}}=\sum\limits_{\sigma \in Vert\mathcal{P}}\mathbf{x}^{%
\mathbf{\sigma }}  \tag{29}
\end{equation}%
so that for the \textit{dilated} polytope the above statement reads as
follows [30,35]: 
\begin{equation}
f(k\mathcal{P},x)=\sum\limits_{\mathbf{m}\in k\mathcal{P}\cap \mathbf{Z}^{d}}%
\mathbf{x}^{\mathbf{m}}=\sum\limits_{i=1}^{n}\mathbf{x}^{\mathbf{kv}%
_{i}}\sum\limits_{\sigma _{i}}\mathbf{x}^{\mathbf{\sigma }_{i}}.  \tag{30}
\end{equation}%
In the last formula the summation is taking place over all vertices whose
location is given by the vectors from the set \{$\mathbf{v}_{1},...,\mathbf{v%
}_{n}\}.$ This means that in actual calculations one can first calculate the
contributions coming from the cones $\sigma _{i}$ of the undilated
(original) polytope $\mathcal{P}$ and only then one can use the last
equation in order to get the result for the dilated polytope.

Let us apply these general results to our specific problem of computation of 
$S(k$ ; $y_{1}$,$y_{2})$ in Eq.(26). We have our simplex with vertices in
x-y plane given by the vector set \{ \textbf{v}$_{1}$=$(0,0)$, \textbf{v}$%
_{2}$=(1,0), \textbf{v}$_{3}$=(0,1)\}, where we have written the x
coordinate first. In this case we have 3 cones: $\sigma _{1}=l_{2}\mathbf{v}%
_{2}+l_{3}\mathbf{v}_{3}$ , $\sigma _{2}=\mathbf{v}_{2}+l_{1}(-$\textbf{v}$%
_{2})+l_{2}($\textbf{v}$_{3}-$\textbf{v}$_{2});\sigma _{3}=$\textbf{v}$%
_{3}+l_{3}($\textbf{v}$_{2}-$\textbf{v}$_{3})+l_{1}(-\mathbf{v}_{3});$\{$%
l_{1}$, $l_{2}$ , $l_{3}$ \}$\in \mathbf{Z}_{\geq 0}$ . In writing these
expressions for the cones we have taken into account that, according to
Brion, when making calculations the apex of each cone should be chosen as
the origin of the coordinate system. Calculation of contributions to the
generating function coming from $\sigma _{1}$ is the most straightforward.
Indeed, in this case we have \textbf{x}=$x_{1}x_{2}=e^{-y_{1}}e^{-y_{2}}.$
Now, the symbol $\mathbf{x}^{\mathbf{\sigma }}$ in Eq.s(29) should be
understood as follows. Since $\sigma _{i}$ , $i=1-3$, is actually a vector,
it has components, like those for \textbf{v}$_{1},$ etc$.$ We shall write
therefore $\mathbf{x}^{\mathbf{\sigma }}=x_{1}^{\sigma (1)}\cdot \cdot \cdot
x_{d}^{\sigma (d)}$ where $\sigma (i)$ is the i-th component of such a
vector. Under these conditions calculation of the contributions from the
first cone with the apex located at (0,0) is completely straightforward and
is given by 
\begin{equation}
\sum\limits_{\left( l_{2},l_{3}\right) \in
Z_{+}^{2}}x_{1}^{l_{2}}x_{2}^{l_{3}}=\frac{1}{1-e_{{}}^{-y_{1}}}\frac{1}{%
1-e^{-y_{2}}}.  \tag{31}
\end{equation}%
It is reduced to the computation of the infinite geometric progression. But
physically, the above result can be looked upon as a product of two
partition functions for two harmonic oscillators whose ground state energy
was discarded. By doing the rest of calculations in the way just described
we reobtain $S(k$ ; $y_{1}$,$y_{2})$ from Eq.(26) as required. This time,
however, we know that the obtained result is associated with the assembly of
harmonic oscillators of frequencies $\pm y_{1}$ and $\pm y_{2}$ and $\pm
(y_{1}-y_{2})$ whose ground state energy is properly adjusted. The
\textquotedblright frequencies\textquotedblright\ (or masses) of these
oscillators are coming from the Morse-theoretic considerations for the
moment maps associated with the critical points of symplectic manifolds as
explained \ in the paper by Atiyah [32]. These masses enter into the
\textquotedblright classical\textquotedblright\ D-H formula, Eq.s(24),(25).
It is just a classical partition function for a system of such described
harmonic oscillators living in the phase space containing critical points.
The D-H classical partition function, Eq.(25), has its quantum analog,
Eq.(26), just described. The ground state for such a quantum system is
degenerate with the degeneracy being described by the Ehrhart polynomial $%
\mathfrak{P}(k,2).$ Such a conclusion is in formal accord with results of
Vergne [13].

Since (by definition) the coefficient of dilation \textit{k=1,2,... , there
is no dynamical system (and its quanum analog) for k=0. But this condition
is the condition for existence of the tachyon pole in the Veneziano
amplitude, Eq.(2). Hence, in view of the results just described this pole
should be considered as unphysical and discarded. }Such arguments are
independent of the analysis made in Ref.[10] where the unphysical tachyons
are removed with help of the properly adjusted phase factors. Clearly, such
factors can be reinstated in the present case as well since their existence
is caused by the requirements of the torus action invariance of the
Veneziano-like amplitudes as explained in [10,14]. Hence, their presence is
consistent with results just presented.

Now we are ready to discuss the Khovanskii-Pukhlikov correspondence. It can
be understood based on the following generic example taken from Ref.[36]. We
would like to compare the integral 
\begin{equation*}
I(z)=\int\limits_{s}^{t}dxe^{zx}=\frac{e^{tz}}{z}-\frac{e^{sz}}{z}\text{ \
with the sum }S(z)=\text{ }\sum\limits_{k=s}^{t}e^{kz}=\frac{e^{tz}}{1-e^{-z}%
}+\frac{e^{sz}}{1-e^{z}}
\end{equation*}
To do so, following Refs[36-38] we introduce the Todd operator (transform)
via 
\begin{equation}
Td(z)=\frac{z}{1-e^{-z}}.  \tag{32}
\end{equation}
Then, it can be demonstrated that 
\begin{equation}
Td(\frac{\partial }{\partial h_{1}})Td(\frac{\partial }{\partial h_{2}}%
)(\int\nolimits_{s-h_{1}}^{t+h_{2}}e^{zx}dx)\mid
_{h_{1}=h_{1}=0}=\sum\limits_{k=s}^{t}e^{kz}.  \tag{33}
\end{equation}
This result can be now broadly generalized. Following Khovanskii and
Pukhlikov [35], we notice that 
\begin{equation}
Td(\frac{\partial }{\partial \mathbf{z}})\exp \left(
\sum\limits_{i=1}^{n}p_{i}z_{i}\right) =Td(p_{1},...,p_{n})\exp \left(
\sum\limits_{i=1}^{n}p_{i}z_{i}\right)  \tag{34}
\end{equation}
By applying this transform to 
\begin{equation}
i(x_{1},...,x_{k};\xi _{1},...,\xi _{k})=\frac{1}{\xi _{1}...\xi _{k}}\exp
(\sum\limits_{i=1}^{k}x_{i}\xi _{i})  \tag{35}
\end{equation}
we obtain, 
\begin{equation}
s(x_{1},...,x_{k};\xi _{1},...,\xi _{k})=\frac{1}{\prod\limits_{i=1}^{k}(1-%
\exp (-\xi _{i}))}\exp (\sum\limits_{i=1}^{k}x_{i}\xi _{i}).  \tag{36}
\end{equation}
This result should be compared now with the individual terms on the r.h.s.
of Eq.(22) on one hand and with the individual terms on the r.h.s of Eq.(23)
on another. Evidently, with help of the Todd transform the exact
''classical'' results for the D-H integral are transformed into the
''quantum'' results of the Brion's identity (22) which is actually
equivalent to the Weyl character formula [28].

We would like to illustrate these general observations by comparing the D-H
result, Eq.(25), with the Weyl character formula result, Eq.(26). To this
purpose we need to use already known data for the cones $\sigma_{i}$ , $%
i=1-3,$ and the convention for the symbol \textbf{x}$^{\sigma}$. In
particular, \ for the first cone we have already $:\mathbf{x}%
^{\sigma_{1}}=x_{1}^{l_{1}}x_{2}^{l_{2}}=\left[ \exp(l_{1}y_{1})\right] \cdot%
\left[ \exp(l_{2}y_{2})\right] \footnote{%
To obtain correct results we needed to change\ signs in front of $l_{1}$ and 
$l_{2}$ . The same should be done for other cones as well.}.$ Now we
assemble the contribution from the first vertex using Eq.(25). We obtain, $%
\left[ \exp(l_1y_1)\right] \cdot\left[ \exp(l_2y_2)\right] /y_1y_2.$ Using
the Todd transform we obtain as well, 
\begin{equation}
Td(\frac{\partial}{\partial l_{1}})Td(\frac{\partial}{\partial l_{2}})\frac {%
1}{y_{1}y_{2}}\left[ \exp(l_{1}y_{1})\right] \cdot\left[ \exp(l_{2}y_{2})%
\right] \mid_{l_{1}=l_{2}=0}=\frac{1}{1-e^{-y_{1}}}\frac{1}{1-e^{-y_{2}}}. 
\tag{37}
\end{equation}
Analogously, for the second cone we obtain: $\mathbf{x}_2^{%
\sigma}=e^{-}ky_1e^{-}l_1y_1e^{-}l_2(y_1-y_2)$ \ so that use of the Todd
transform produces: 
\begin{equation}
Td(\frac{\partial}{\partial l_{1}})Td(\frac{\partial}{\partial l_{2}})\frac {%
1}{y_{1}\left( y_{1}-y_{2}\right) }%
e^{-ky_{1}}e^{-l_{1}y_{1}}e^{-l_{2}(y_{1}-y_{2})}\mid_{l_{1}=l_{2}=0}=\frac{1%
}{1-e^{y_{1}}}\frac{e^{-ky_{1}}}{1-e^{y_{1}-y_{2}}},  \tag{38}
\end{equation}
etc.

The obtained results can now be broadly generalized. To this purpose we can
formally rewrite the partition function, Eq.(24), in the following symbolic
form 
\begin{equation}
I(k,\mathbf{f})=\int\limits_{k\Delta }d\mathbf{x}\exp \mathbf{(-f\cdot x)} 
\tag{39}
\end{equation}
valid for any finite dimension $d$. Since we have performed all calculations
explicitly for two dimensional case, for the sake of space, we only provide
the idea behind such type of calculation\footnote{%
Mathematically inclined reader is encoraged to read paper by Brion and
Vergne, Ref.[39], where all missing details are scrupulously presented.}. In
particular, using (25) we can rewrite this integral formally as follows 
\begin{equation}
\int\limits_{k\Delta }d\mathbf{x}\exp \mathbf{(-f\cdot x)=}\sum\limits_{p}%
\frac{\exp (-\mathbf{f}\cdot \mathbf{x(}p\mathbf{)})}{\prod%
\limits_{i}^{d}h_{i}^{p}(\mathbf{f})}  \tag{40}
\end{equation}
Applying the Todd operator (transform) to both sides of this formal
expression and taking into account Eq.s(37),(38) (providing assurance that
such an operation indeed is legitimate and makes sense) we obtain, 
\begin{align}
\int\limits_{k\Delta }d\mathbf{x}\prod\limits_{i=1}^{d}\frac{x_{i}}{1-\exp
(-x_{i})}\exp \mathbf{(-f\cdot x)}& =\sum\limits_{\mathbf{v}\in Vert\mathcal{%
P}}\exp \{<\mathbf{f}\cdot \mathbf{v}>\}\left[ \prod\limits_{i=1}^{d}(1-\exp
\{-h_{i}^{v}(\mathbf{f})u_{i}^{v}\})\right] ^{-1}  \notag \\
& =\sum\limits_{\mathbf{x}\in \mathcal{P\cap }\mathbf{Z}^{d}}\exp \{<\mathbf{%
f}\cdot \mathbf{x}>\}  \tag{41}
\end{align}
where the last line was written in view of Eq.(22). From here, in the limit
: $\mathbf{f}=0$ we reobtain $p(k,n)$ defined in Eq.(7). Thus, using
classical partition function, Eq.(39), (discussed in the form of Exercises
2.27 and 2.28 \ in the book, Ref. [37], by Guillemin) and applying to it the
Todd transform we recover the quantum mechanical partition function whose
ground state provides us with the combinatorial factor $p(k,n).$

\section{\protect\bigskip From analysis to synthesis}

\subsection{\protect\bigskip The Poincare$^{\prime }$ polynomial}

The results discussed earlier are obtained for some fixed dilation factor $k$
. In view of \ (6), \ they can be \ rewritten in the form valid for any
dilation factor $k$. To this purpose it is convenient to rewrite (6) in the
following equivalent form: 
\begin{align}
\frac{1}{\det (1-Mt)}& =\frac{1}{(1-tz_{0})\cdot \cdot \cdot (1-tz_{k})}%
=(1+tz_{0}+\left( tz_{0}\right) ^{2}+...)\cdot \cdot \cdot (1+tz_{n}+\left(
tz_{n}\right) ^{2}+...)  \notag \\
& =\sum\limits_{n=0}^{\infty
}(\sum\limits_{k_{0}+...+k_{k}=n}z_{0}^{k_{0}}\cdot \cdot \cdot
z_{k}^{k_{k}})t^{n}\equiv \sum\limits_{n=0}^{\infty }tr(M_{n})t^{n}, 
\tag{42}
\end{align}
where the linear map from $k+1$ dimensional vector space $V$ to $V$ is given
by $\ $matrix $M\in G\subset GL(V)$ whose eigenvalues are $z_{0},...,z_{k}.$
Using this observation several conclusions can be drawn. First, it should be
clear that 
\begin{equation}
\sum\limits_{k_{0}+...+k_{k}=n}z_{0}^{k_{0}}\cdot \cdot \cdot
z_{k}^{k_{k}}=\sum\limits_{\mathbf{m}\in n\Delta \cap \mathbf{Z}^{k+1}}%
\mathbf{x}^{\mathbf{m}}=tr(M_{n}).  \tag{43}
\end{equation}
Second, following Stanley [3,15] we would like to consider the algebra of
invariants of $G$. To this purpose we introduce a basis $\mathbf{x}=$\{ $%
x_{0},...,x_{k}\}$ of $V$ and the polynomial ring $R=\mathbf{C}$[$%
x_{0},...,x_{k}]$ so that if $f\in R$ , then $Mf(\mathbf{x})=f(M\mathbf{x})$%
. The algebra of invariant polynomials $R^{G}$ can be defined now as 
\begin{equation*}
R^{G}=\{f\in R:Mf(\mathbf{x})=f(M\mathbf{x})=f(\mathbf{x})\text{ \ }\forall
M\in G\}.
\end{equation*}
These invariant polynomials can be explicitly constructed as averages over
the group $G$ according to prescription: 
\begin{equation}
Av_{G}f=\frac{1}{\left| G\right| }\sum\limits_{M\in G}Mf,  \tag{44}
\end{equation}
with $\left| G\right| $ being the cardinality of $G$. Suppose now that $f\in
R^{G}$, then, evidently, $f\in R^{G}=Av_{G}f$ so that $Av_{G}^{2}f=Av_{G}f=f$
. Hence, the operator $Av_{G}$ is indepotent. Because of this, its
eigenvalues can be only 1 and 0. From here it follows that 
\begin{equation}
\dim f_{n}^{G}=\frac{1}{\left| G\right| }\sum\limits_{M\in G}tr(M_{n}). 
\tag{45}
\end{equation}
Thus far our analysis was completely general. To obtain Eq.(7) we have to
put $z_{0}=...=z_{k}=1$ in (6). This time, however we can use the obtained
results in order to write the following expansion for the Poincar$e^{\prime
} $ polynomial [3,15,28] which for the appropriately chosen $G$ is
equivalent to (8): 
\begin{equation}
P(R^{G},t)=\sum\limits_{n=0}^{\infty }\frac{1}{\left| G\right| }%
\sum\limits_{M\in G}tr(M_{n})t^{n}=\sum\limits_{n=0}^{\infty }\dim
f_{n}^{G}t^{n}.  \tag{46}
\end{equation}
Evidently, the Ehrhart polynomial $\mathfrak{P}(k,n)=\dim f_{n}^{G}.$ To
figure out the group $G$ in the present case is easy since, actually, \ the
group is trivial: $G=1.$This is so because the eigenvalues $z_{0}$, ..., $%
z_{k}$ of the matrix $M$ all are equal to 1. It should be clear, however,
that for some appropriately chosen group $G$ expansion (17) is also the
Poincare polynomial (for the Cohen -Macaulay polynomial algebra [3,15,28]).
This fact provides independent (of Refs. [9,10]) evidence that both the
Veneziano and Veneziano-like amplitudes are of topological origin.

\subsection{Connections with intersection theory}

We would like to strengthen this observation now. To this purpose, in view
of (23), and taking into account that for the symplectic 2-form $\Omega
=\sum\nolimits_{i=1}^{k}dx_{i}\wedge dy_{i}$ the $n$-th power is given by $%
\Omega ^{n}=\Omega \wedge \Omega \wedge \cdot \cdot \cdot \wedge \Omega $ =$%
dx_{1}\wedge dy_{1}\wedge \cdot \cdot \cdot dx_{n}\wedge dy_{n}$, it is
convenient to introduce the differential form 
\begin{equation}
\exp \Omega =1+\Omega +\frac{1}{2!}\Omega \wedge \Omega +\frac{1}{3!}\Omega
\wedge \Omega \wedge \Omega +\cdot \cdot \cdot \text{ \ \ \ .}  \tag{47}
\end{equation}%
By design, the expansion \ in the r.h.s. will have only $k$ terms. The form $%
\Omega $ is closed, i.e. $d\Omega =0$ (the Liouvolle theorem), but not
exact. In view of the expansion (47) the D-H integral (39) can be rewritten
as 
\begin{equation}
I(k,\mathbf{f})=\int\limits_{k\Delta }\exp \mathbf{(}\tilde{\Omega}\mathbf{)}
\tag{48}
\end{equation}%
where, following Atiyah and Bott [16] we have introduced the form $\tilde{%
\Omega}=\Omega -\mathbf{f\cdot x.}$ Doing so requires to replace the
exterior derivative $d$ acting on $\Omega $ by $\tilde{d}=d+i(\mathbf{x})$ \
(where the operator $i(\mathbf{x})$ reduces the degree of the form by one)
with respect to which the form $\tilde{\Omega}$ is equivariantly closed, i.e.%
$\tilde{d}\tilde{\Omega}=0.$ More explicitly, we have $\tilde{d}\tilde{\Omega%
}=d\Omega +i(\mathbf{x})\Omega -\mathbf{f\cdot dx=}0$. Since $d\Omega =0,$
we obtain the equation for the moment map : $i(\mathbf{x})\Omega -\mathbf{%
f\cdot dx=}0$ [31,37]. If use of the operator $d$ on differential forms
leads to the notion of cohomology, use of the operator $\tilde{d}$ leads to
the notion of equivariant cohomology. Although details can be found in the
paper by Atiyah and Bott [16], more relaxed pedagogical exposition can be
found in the monograph by Guillemin and Sternberg [40]. To make further
progress, we would like to rewrite the two-form $\Omega $ in complex
notations [31]. To this purpose, we introduce $z_{j}=p_{j}+iq_{j}$ and its
complex conjugate. In terms of these variables $\Omega $ acquires the
following form $:\Omega =$ $\frac{i}{2}\sum\nolimits_{i=1}^{k}dz_{i}\wedge d%
\bar{z}_{i}.$ Next, recall [18] that for any K\"{a}hler manifold the
fundamental 2-form $\Omega $ can be written as $\Omega =\frac{i}{2}%
\sum\nolimits_{ij}h_{ij}(z)dz_{i}\wedge d\bar{z}_{j}$ \ provided that $%
h_{ij}(z)=\delta _{ij}+O(\left\vert z\right\vert ^{2})$. This means that in
fact all K\"{a}hler manifolds are symplectic [14,34]. On such K\"{a}hler
manifolds one can introduce the Chern cutrvature 2-form which (up to a
constant ) should look like $\Omega .$ \ It should belong to the first Chern
class [19]. This means that, at least formally, consistency reqiures us to
identify x$_{i}^{\prime }s$ entering the product $\mathbf{f\cdot x}$ in the
form $\tilde{\Omega}$ with the first Chern classes $c_{i}$, i.e. $\mathbf{%
f\cdot x\equiv }\sum\nolimits_{i=1}^{d}f_{i}c_{i}.$ This fact was proven
rigorously in the above mentioned paper by Atiyah and Bott [16]. Since in
the Introduction we already mentioned that the Veneziano amplitudes can be
formally associated with the period integrals for the Fermat (hyper)surfaces 
$\mathcal{F}$ and since such integrals can be interpreted as intersection
numbers between the cycles on $\mathcal{F}$ [16,19] (see also Ref.[37],
page72) one can formally rewrite the \textit{precursor} to the Veneziano
amplitude [10] as 
\begin{equation}
I=\left( \frac{-\partial }{\partial f_{0}}\right) ^{r_{0}}\cdot \cdot \cdot
\left( \frac{-\partial }{\partial f_{d}}\right) ^{r_{d}}\int\limits_{\Delta
}\exp (\tilde{\Omega})\mid _{f_{i}=0\text{ }\forall i}=\int\limits_{\Delta }d%
\mathbf{x(}c_{0})^{r_{0}}\cdot \cdot \cdot (c_{d})^{r_{d}}  \tag{49}
\end{equation}%
provided that $r_{0}+\cdot \cdot \cdot +r_{d}=n$ \ in in view of Eq.(11).
Analytical continuation of such an integral (as in the case of usual beta
function) then will produce the Veneziano amplitudes. In such a language,
calculation of the Veneziano amplitudes using generating function, Eq.(48),
mathematically becomes almost equivalent to calculations of averages in the
Witten-Kontsevich model [20-22]. In \ addition, as was also noticed by
Atiyah and Bott [16], the replacement of the exterior derivative $d$ by $%
\tilde{d}=d+i(\mathbf{x})$ was inspired by earlier work by Witten on
supersymmetric formulation of quantum mechanics and Morse theory [17]. Such
an observation along with results of Ref.[40] allows us to develop
calculations of the Veneziano amplitudes using supersymmetric formalism.

\subsection{Supersymmetry and the Lefshetz isomomorphism}

We begin with the following observations. Let $X$ be the complex Hermitian
manifold and let $\mathcal{E}^{p+q}(X)$ denote the complex -valued
differential forms (sections) of type $(p,q)$ $,p+q=r,$ living on $X$. The
Hodge decomposition insures that $\mathcal{E}^{r}(X)$=$\sum\nolimits_{p+q=r}%
\mathcal{E}^{p+q}(X).$ The Dolbeault operators $\partial $ and $\bar{\partial%
}$ act on $\mathcal{E}^{p+q}(X)$ according to the rule \ $\partial :\mathcal{%
E}^{p+q}(X)\rightarrow \mathcal{E}^{p+1,q}(X)$ and $\bar{\partial}:\mathcal{E%
}^{p+q}(X)\rightarrow \mathcal{E}^{p,q+1}(X)$ , so that the exterior
derivative operator is defined as $d=\partial +\bar{\partial}$. Let now $%
\varphi _{p}$,$\psi _{p}\in \mathcal{E}^{p}$. By analogy with traditional
quantum mechanics we define (using Dirac's notations) the inner product 
\begin{equation}
<\varphi _{p}\mid \psi _{p}>=\int\limits_{M}\varphi _{p}\wedge \ast \bar{\psi%
}_{p}  \tag{50}
\end{equation}
where the bar means the complex conjugation and the star $\ast $ means the
usual Hodge conjugation. Use of such a product is motivated by the fact that
the period integrals, e.g. those for the Veneziano-like amplitudes, and,
hence, those given by Eq.(49), are expressible through such inner products
[19]. Fortunately, such a product possesses properties typical for the
finite dimensional quantum mechanical Hilbert spaces. In particular, 
\begin{equation}
<\varphi _{p}\mid \psi _{q}>=C\delta _{p,q}\text{ and }<\varphi _{p}\mid
\varphi _{p}>>0,  \tag{51}
\end{equation}
where $C$ is some positive constant. With respect to such defined scalar
product it is possible to define all conjugate operators, e.g. $d^{\ast }$,
etc. and, most importantly, the Laplacians 
\begin{align}
\Delta & =dd^{\ast }+d^{\ast }d,  \notag \\
\square & =\partial \partial ^{\ast }+\partial ^{\ast }\partial ,  \tag{52}
\\
\bar{\square}& =\bar{\partial}\bar{\partial}^{\ast }+\bar{\partial}^{\ast }%
\bar{\partial}.  \notag
\end{align}
All this was known to mathematicians before Witten's work [17\textbf{].} The
unexpected twist occurred when Witten suggested to extend the notion of the
exterior derivative $d$. Within the de Rham picture (valid for both real and
complex manifolds) let $M$ be a compact Riemannian manifold and $K$ be the
Killing vector field which is just one of the generators of isometry of $M,$
then Witten suggested to replace the exterior derivative operator $d$ by the
extended operator 
\begin{equation}
d_{s}=d+si(K)  \tag{53}
\end{equation}
briefly discussed earlier in the context of the equivariant cohomology. Here 
$s$ is real nonzero parameter conveniently chosen. Witten argues that one
can construct the Laplacian (the Hamiltonian in his formulation) $\Delta $
by replacing $\Delta $ by $\Delta _{s}=d_{s}d_{s}^{\ast }+d_{s}^{\ast }d_{s}$
. This is possible if and only if $d_{s}^{2}=d_{s}^{\ast 2}$ $=0$ or, since $%
d_{s}^{2}=s\mathcal{L}(K)$ , where $\mathcal{L}(K)$ is the Lie derivative
along the field $K$, if the Lie derivative acting on the corresponding
differential form vanishes. The details are beautifully explained in the
much earlier paper by Frankel [41]. Atiyah and Bott\ observed that the
auxiliary parameter \textbf{s} can be identified with earlier introduced 
\textbf{f}. This observation provides the link between the D-H formalism
discussed earlier and Witten's supersymmetric quantum mechanics.

Looking at Eq.s (52) and following Ref.s[13\textbf{,}31\textbf{,}37] we
consider the (Dirac) operator $\acute{\partial}=\bar{\partial}+\bar{\partial}%
^{\ast }$ and its adjoint with respect to scalar product, Eq.(50). Then use
of above references suggests that the dimension $Q$ of the quatum Hilbert
space associated with the reduced phase space of the D-H integral considered
earlier is given by 
\begin{equation}
Q=\ker \acute{\partial}-co\ker \acute{\partial}^{\ast }.  \tag{54}
\end{equation}%
Such a definition was\ also used by Vergne[13]. In view of the results of
the previous section, and, in accord with Ref.[13], we make an
identification: $Q=\mathfrak{P}(k,n).$

We would like to arrive at this result using different set of arguments. To
this purpose we notice first that according to Theorem 4.7. by Wells [18] we
have $\Delta =2\square =2\bar{\square}$ with respect to the K\"{a}hler
metric on $X$. Next, according to the Corollary 4.11. of the same reference $%
\Delta $ commutes with $d,d^{\ast },\partial ,\partial ^{\ast },\bar{\partial%
}$ and $\bar{\partial}^{\ast }.$ From these facts it follows immediately
that if we, in accord with Witten, choose $\Delta $ as our Hamiltonian, then
the supercharges can be selected as $Q^{+}=d+d^{\ast }$ and $Q^{-}=i\left(
d-d^{\ast }\right) .$ Evidently, this is not the only choice as Witten also
indicates. If the Hamiltonian \ H is acting in \textit{finite} dimensional
Hilbert space one may require axiomatically that : a) there is a vacuum
state (or states) $\mid \alpha >$ such that H$\mid \alpha >=0$ (i.e. this
state is the harmonic differential form) and $Q^{+}\mid \alpha >=Q^{-}\mid
\alpha >=0$ . This implies, of course, that [H,$Q^{+}]=[$H,$Q^{-}]=0.$
Finally, \ once again, following Witten, we may require that $\left(
Q^{+}\right) ^{2}=\left( Q^{-}\right) ^{2}=$H. Then, the equivariant
extension, Eq.(53), leads to $\left( Q_{s}^{+}\right) ^{2}=$ H+$2is\mathcal{L%
}$($K$). Fortunately, the above \ supersymmetry algebra can be extended. As
it is mentioned in Ref.[19], there are operators acting on differential
forms living on K\"{a}hler (or Hodge) manifolds whose commutators are
isomorphic to $sl_{2}(\mathbf{C})$ Lie algebra. It is known [42] that 
\textit{all} semisimple Lie algebras are made of copies of $sl_{2}(\mathbf{C}%
)$. Now we can exploit these observations using the Lefschetz isomorphism
theorem whose exact formulation is given as Theorem 3.12 in the book by
Wells, Ref. [19]. We are only using some parts of this theorem in our work.

In particular, using notations of this reference we introduce the operator $%
L $ commuting with $\Delta $ and its adjoint $L^{\ast }\equiv \Lambda $ .\
It can be shown, Ref. [19], page 159, that $L^{\ast }=w\ast L\ast $ where,
as before, $\ast $ denotes the Hodge star operator and the operator $w$ can
be formally defined through the relation $\ast \ast =w$, Ref.[19] page 156.
From these definitions it should be clear that $L^{\ast }$ also commutes
with $\Delta $ on the space of harmonic differential forms (in accord with
page 195 of [19]). As part of the preparation for proving of the Lefschetz
isomorphism theorem, it can be shown [19], that 
\begin{equation}
\lbrack \Lambda ,L]=B\text{ and }[B,\Lambda ]=2\Lambda \text{, }[B,L]=-2L. 
\tag{55}
\end{equation}
At the same time, the Jacobson-Morozov theorem, Ref.[24], and \ results of
Ref.[42], page 37, essentially guarantee that any $sl_{2}(\mathbf{C})$ Lie
algebra can be brought into form 
\begin{equation}
\lbrack h_{\alpha },e_{\alpha }]=2e_{\alpha }\text{ , }[h_{\alpha
},f_{\alpha }]=-2f_{\alpha }\text{ , \ }[e_{\alpha },f_{\alpha }]=h_{\alpha
}\   \tag{56}
\end{equation}
upon appropriate rescaling. The index $\alpha $ counts number of $sl_{2}(%
\mathbf{C})$ algebras in a semisimple Lie algebra. Comparison between the
above two expressions leads to the Lie algebra endomorphism, i.e. the
operators $h_{\alpha },f_{\alpha }$ and $e_{\alpha }$ act on the vector
space $\{v\}$ to be described below while the operators $\Lambda ,L$ and $B$
obeying the same commutation relations act on the space of differential
forms. It is possible to bring Eq.s(55) and (56) to even closer
correspondence. To this purpose, following Dixmier [43], Ch-r 8, we
introduce operators $h=\sum\nolimits_{\alpha }a_{\alpha }h_{\alpha }$, $%
e=\sum\nolimits_{\alpha }b_{\alpha }e_{\alpha }$, $f=\sum\nolimits_{\alpha
}c_{\alpha }f_{\alpha }.$ Then, provided that the constants are subject to
constraint: $b_{\alpha }c_{\alpha }=a_{\alpha }$ , the commutation relations
between the operators $h$, $e$ and $f$ are \textit{exactly the same} as for $%
B$, $\Lambda $ and $L$ respectively. To avoid unnecessary complications, we
choose $a_{\alpha }=b_{\alpha }=c_{\alpha }=1$.

Next, following Serre, Ref. [23], Ch-r 4, we need to introduce the notion of
the \textit{primitive} vector (or element).This is the vector $v$ such that $%
hv$=$\lambda v$ but $ev=0.$ The number $\lambda $ is the weight of the
module $V^{\lambda }=\{v\in V\mid hv$=$\lambda v\}.$ If the vector space is 
\textit{finite dimensional}, then $V=\sum\nolimits_{\lambda }V^{\lambda }$ .
Moreover, only if $V^{\lambda }$ is finite dimensional it is straightforward
to prove that the primitive element does exist. The proof is based on the
observation that if $x$ is the eigenvector of $h$ with weight $\lambda ,$
then $ex$ is also the eigenvector of $h$ with eigenvalue $\lambda -2,$ etc.
Moreover, from the book by Kac [44], Chr.3, it follows that if $\lambda $ is
the weight of $V,$ then $\lambda -<\lambda ,\alpha _{i}^{\vee }>\alpha _{i}$
is also the weight with the same multiplicity, provided that $<\lambda
,\alpha _{i}^{\vee }>\in \mathbf{Z}$\textbf{\ \ . }Kac therefore introduces
another module: $U=\sum\nolimits_{k\in \mathbf{Z}}$ $V^{\lambda +k\alpha
_{i}}$ \ . Such a module is finite for finite Weyl-Coxeter reflection groups
and is infinite for the affine reflection groups associated with the affine
Kac-Moody Lie algebras.

We would like to argue that for our purposes it is sufficient to use only 
\textit{finite} reflection (or pseudo-reflection) groups. It should be
clear, however, from reading the book by Kac that the infinite dimensional
version of the module $U$ leads straightforwardly to all known
string-theoretic results. In the case of CFT this is essential, but for
calculation of the Veneziano-like amplitudes this is \textit{not} essential
as we are about to demonstrate.Indeed, by accepting \ the traditional option
we\ \ loose at once our connections with the Lefschetz isomorphism theorem (
relying heavily on the existence of primitive elements) and with the Hodge
theory in its standard form on which our arguments are based. The infinite
dimensional extensions of the Hodge-de Rham theory involving loop groups,
etc. relevant for CFT can be found in Ref.[45]. Fortunately, they are not
needed for our calculations. Hence, below we work only with the finite
dimensional spaces.

In particular, let $v$ be a primitive element of weight $\lambda $ then,
following Serre, we let $v_{n}=\frac{1}{n!}e^{n}v$ for $n\geq 0$ and $%
v_{-1}=0,$ so that 
\begin{align}
hv_{n}& =(\lambda -2n)v_{n}  \tag{57} \\
ev_{n}& =(n+1)v_{n+1}  \notag \\
fv_{n}& =(\lambda -n+1)v_{n-1}.  \notag
\end{align}%
Clearly, the operators $e$ and $f$ are the creation and the annihilation
operators according to the existing in physics terminology while the vector $%
v$ can be interpreted as the vacuum state vector. The question arises: how
this vector is related to the earlier introduced vector $\mid \alpha >?$
Before providing an answer to this question we need, following Serre, to
settle the related issue. In particular, we can either: a) assume that for
all $n\geq 0$ the first of Eq.s(57) has solutions and all vectors $%
v,v_{1},v_{2}$ , ...., are linearly independent or b) beginning from some $%
m+1\geq 0,$ all vectors $v_{n\text{ }}$are zero, i.e. $v_{m}\neq 0$ but $%
v_{m+1}=0.$ The first option leads to the infinite dimensional
representations associated with Kac-Moody affine algebras just mentioned.
The second option leads to the finite dimensional representations and to the
requirement $\lambda =m$ with $m$ being an integer. Following Serre, this
observation can be exploited further thus leading us to crucial physical
identifications. Serre observes that with respect to $n=0$ Eq.s(57) possess
a (\textquotedblright super\textquotedblright )symmetry. That is the linear
mappings 
\begin{equation}
e^{m}:V^{m}\rightarrow V^{-m}\text{ and \ }f^{m}:V^{-m}\rightarrow V^{m} 
\tag{58}
\end{equation}%
are isomorphisms and the dimensionality of $V^{m}$ and $V^{-m}$ are the
same. Serre provides an operator (the analog of Witten's $F$ operator) $%
\theta =\exp (f)\exp (e)\exp (-f)$ such that $\theta \cdot f=-e\cdot \theta $%
, $\theta \cdot e=-\theta \cdot f$ and $\theta \cdot h=-h\cdot \theta .$ In
view of such an operator, it is convenient to redefine $h$ operator : $%
h\rightarrow \hat{h}=h-\lambda $. Then, for such redefined operator the
vacuum state is just $v$. Since both $L$ and $L^{\ast }=\Lambda $ commute
with the supersymmetric Hamiltonian H and, because of the group
endomorphism, we conclude that the vacuum state $\mid \alpha >$ for H
corresponds to the primitive state vector $v$.

Now we are ready to apply yet another isomorphism following Ginzburg [24,
Chap. 4, pages 205-206] \footnote{%
Unfortunately, the original sourse contains minor mistakes. These are easily
correctable. The corrected results are given in the text.}. To this purpose
we make the following identification 
\begin{equation}
e_{i}\rightarrow t_{i+1}\frac{\partial }{\partial t_{i}}\text{ , }%
f_{i}\rightarrow t_{i}\frac{\partial }{\partial t_{i+1}}\text{ , }%
h_{i}\rightarrow 2\left( t_{i+1}\frac{\partial }{\partial t_{i+1}}-t_{i}%
\frac{\partial }{\partial t_{i}}\right) ,.  \tag{59}
\end{equation}%
$i=0,...,m.$ Such operators are acting on the vector space made of monomials
of the type 
\begin{equation}
v_{n}\rightarrow \mathcal{F}_{n}=\frac{1}{n_{0}!n_{1}!\cdot \cdot \cdot
n_{k}!}t_{0}^{n_{0}}\cdot \cdot \cdot t_{k}^{n_{k}}  \tag{60}
\end{equation}%
where $n_{0}+...+n_{k}=n$ . This result is useful to compare with Eq. (49).

Eq.s (57) have now their analogs 
\begin{align}
h_{i}\ast \mathcal{F}_{n}(i)& =2(n_{i+1}-n_{i})\mathcal{F}_{n}(i)  \notag \\
e_{i}\ast \mathcal{F}_{n}(i)& =2n_{i}\mathcal{F}_{n}(i+1)  \tag{61} \\
\text{ }f_{i}\ast \mathcal{F}_{n}(i)& =2n_{i+1}\mathcal{F}_{n}(i-1),  \notag
\end{align}%
where, clearly, one should make the following consistent identifications: $%
m(i)-2n(i)=2\left( n_{i+1}-n_{i}\right) $ , $2n_{i}=n(i)+1$ and $%
m(i)-n(i)+1=2n_{i+1}.$ Next, we define the total Hamiltonian: $h=$ $%
\sum\nolimits_{i=0}^{k}h_{i}$ so that$\sum\nolimits_{i=0}^{k}m(i)=n,$ and
then consider its action on one of the wave functions of the type given by
Eq.(60).Since the operators defined by Eq.s (59) by design preserve the
total degree of monomials of the type given by Eq.(60) (that is they
preserve the Veneziano energy-momentum codition), we obtain the ground state
degeneracy equal to $\mathfrak{P}(k,n)$ in agreement with Vergne, Ref. [13],
where it was obtained using different\ methods. \ Clearly, the factor $%
\mathfrak{P}(k,n)$ is just the number of solutions in nonnegative integers
to $n_{0}+...+n_{k}=n$, Ref.[33], page \ 252.

\bigskip

\pagebreak

\bigskip

\textbf{References}

\bigskip

[1] \ G.Veneziano, \ Construction of crossing symmetric, Regge

\ \ \ \ \ behaved, amplitude for linearly rising trajectories,

\ \ \ \ \ \textit{Il Nuovo Chimento} \textbf{57}A (1968) 190-197.

[2] \ M.Green, J.Schwarz, E.Witten, \textit{Superstring Theory}, vol.1,

\ \ \ \ \ (Cambridge U.Press, Cambridge, UK, 1987).

[3] \ R.Stanley, \textit{Combinatorics and Commutative Algebra}

\ \ \ \ \ \ (Birkh\"{a}user, Boston, MA, 1996).

[4] \ S.Chowla,\ A.Selberg, \ On Epstein's Zeta function,

\ \ \ \ \ \textit{J. Reine Angew.Math.} \textbf{227} (1967) 86-100.

[5] \ A.Weil, Abelian varieties and Hodge ring, \textit{Collected Works},

\ \ \ \ \ vol.3 (Springer-Verlag, Berlin, 1979).

[6] \ A.Weil, Sur les periods des integrales Abeliennes,

\ \ \ \ \textit{Comm.Pure Appl.\ Math}. \textbf{29} (1976) 81-819.

[7] \ B.Gross, On periods of Abelian integrals and formula of

\ \ \ \ Chowla and Selberg, \textit{Inv.Math}.\textbf{45} (1978) 193-211.

[8] \ S.Lang, \textit{Introduction to Algebraic and Abelian Functions}

\ \ \ \ \ (Springer-Verlag, Berlin, 1982).

[9] \ A.Kholodenko, New string amplitudes from old Fermat

\ \ \ \ \ (hyper)surfaces, \textit{IJMP} A\textbf{19} (2004) 1655-1703.

[10] A.Kholodenko, New strings for old Veneziano amplitudes I.

\ \ \ \ \ \ Analytical treatment, \textit{J.Geom.Phys}. (2005) in press;

\ \ \ \ \ \ arXiv: hep-th/040242.

[11] P.Deligne, Hodge cycles and Abelian varieties,

\ \ \ \ \ \ \textit{LNM}. \textbf{900} (1982) 9-100.

[12] J.Carlson, S.Muller-Stach, C.Peters, \textit{Period Mapping}

\ \ \ \ \ \ \textit{and Period Domains} (Cambridge U.Press, Cambridge, UK,
2003).

[13] M.Vergne, Convex polytopes and quantization of symplectic manifolds,

\ \ \ \ \ \textit{PNAS} \textbf{93} (1996) 14238-14242.

[14] M.Audin, \textit{Torus Actions on Symplectic Manifolds}

\ \ \ \ \ \ (Birkh\"{a}user, Boston, MA, 2004).

[15] R.Stanley, Invariants of finite groups and their applications to

\ \ \ \ \ \ combinatorics, \textit{BAMS} \textbf{1} (1979) 475-511.

[16] M.Atiyah, R.Bott, The moment map and equivariant cohomology,

\ \ \ \ \ \ \textit{Topology }\textbf{23} (1984) 1-28.

[17] E.Witten, Supersymmetry and Morse theory,

\ \ \ \ \ \ \textit{J.Diff.Geom}. \textbf{17} (1982) 661-692.

[18] W.Fulton, \textit{Introduction to Toric Varieties}

\ \ \ \ \ \ \ (Princeton U. Press, Princeton, 1993).

[19] R.Wells, \textit{Differential Analysis on Complex Manifolds}

\ \ \ \ \ \ \ \ (Springer-Verlag, Berlin, 1980).

[20] E.Witten, Two dimensional gravity and intersection theory on

\ \ \ \ \ \ moduli space, \textit{Surv.Diff.Geom}.\textbf{1} (1991) 243-310.

[21] M.Kontsevich, Intersection theory on the moduli space of curves

\ \ \ \ \ \ and the matrix Airy function, \textit{Comm.Math.Phys}. \textbf{%
147 }(1992) 1-23.

[22] A.Kholodenko, Kontsevich-Witten model from 2+1 gravity: new exact

\ \ \ \ \ \ combinatorial solution, \textit{J.Geom.Phys}. \textbf{43} (2002)
45-91.

[23] J-P.Serre, \textit{Algebres de Lie Simi-Simples Complexes},

\ \ \ \ \ \ (Benjamin , Inc. New York, 1966).

[24] V.Ginzburg, \textit{Representation Theory and Complex Geometry}

\ \ \ \ \ \ (Birkh\"{a}user-Verlag, Boston, 1997).

[25] F.Hirzebruch, D.Zagier, \textit{The Atiyah-Singer Theorem and Elementary%
}

\ \ \ \ \ \ \textit{Number Theory} (Publish or Perish, Berkeley, CA , 1974).

[26] V.Buchstaber, T.Panov, \textit{Torus actions and Their}

\ \ \ \ \ \ \ \textit{Applications in Topology and Combinatorics}

\ \ \ \ \ \ (AMS Publishers, Providence, RI, 2002).

[27] R.Dias, S.Robins, \ The Ehrhart polynomial of a lattice polytope,

\ \ \ \ \ \textit{\ Ann.Math.} \textbf{145} (1997) 503-518.

[28] R.Kane, \textit{Reflection Grups and Invariant Theory}

\ \ \ \ \ \ (Springer-Verlag, Berlin, 2001).

[29] A.Barvinok, Computing the volume, counting integral points,

\ \ \ \ \ and exponential sums, \textit{Discr.Comp.Geometry} \textbf{10}
(1993) 123-141.

[30] M.Brion, Points entiers dans les polyedres convexes,

\ \ \ \ \ \ \textit{Ann.Sci.Ecole Norm. Sup}. \textbf{21} (1988) 653-663.

[31] V.Guillemin, V.Ginzburg, Y.Karshon, \textit{Moment Maps,}

\ \ \ \ \ \ \textit{Cobordisms, and Hamiltonian Group Actions}

\ \ \ \ \ \ (AMS Publishers, Providence, RI, 2002).

[32] M.Atiyah, \ Convexity and commuting Hamiltonians,

\ \ \ \ \ \ \textit{London Math.Soc.Bull}. \textbf{14} (1982)1-15.

[33] M.Vergne, In E.Mezetti, S.Paycha (Eds),

\ \ \ \ \ \ \textit{European Women in Mathematics}, pp 225-284,

\ \ \ \ \ \ (World Scientific, Singapore, 2003).

[34] M.Atiyah, Angular \ momentum, convex polyhedra

\ \ \ \ \ \ and algebraic geometry,

\ \ \ \ \ \ \textit{Proc. Edinburg Math.Soc}. \textbf{26} (1983)121-138.

[35] A.Barvinok, \textit{A Course in Convexity},

\ \ \ \ \ \ \ (AMS Publishers, Providence, RI, 2002).

[36] M.Brion, M.Vergne, Lattice points in simple polytopes,

\ \ \ \ \ \textit{J.AMS} \textbf{10} (1997) 371-392.

[37] V.Guillemin, \textit{Moment Maps and Combinatorial}

\ \ \ \ \ \ \textit{Invariants of Hamiltonian T}$^{n}$\textit{\ Spaces}

\ \ \ \ \ \ (Birkh\"{a}user, Boston, MA, 1994).

[38] A.Khovanskii, A.Pukhlikov, A Riemann--Roch theorem

\ \ \ \ \ for integrals and sums of quasipolynomials over virtual

\ \ \ \ \ \ polytopes, \textit{St.Petersburg Math.J}. \textbf{4} (1992)
789-812.

[39] M.Brion, M.Vergne, \ An equivariant Riemann-Roch theorem

\ \ \ \ \ for complete simplicial toric varieties,

\ \ \ \ \ \textit{\ J.Reine Angew.Math}. \textbf{482} (1997) 67-92.

[40] V.Guillemin, S.Sternberg, \textit{Supersymmetry and Equivariant}

\ \ \ \ \ \ \textit{de Rham Theory} ( Springer-Verlag, Berlin, 1999).

[41] T.Frankel, Fixed points and torsion on K\"{a}hler manifolds,

\ \ \ \ \ \ Ann.Math.\textbf{70} (1959) 1-8.

[42] J.Humphreys, \textit{Introduction to Lie Algebras and Representations}

\ \ \ \ \ \ \textit{Theory} (Springer-Verlag, Berlin, 1972).

[43] J.Dixmier, \textit{Enveloping Algebras}

\ \ \ \ \ \ (Elsevier, Amsterdam, 1977).

[44] V.Kac, \textit{Infinite Dimensional Lie Algebras}

\ \ \ \ \ \ (Cambridge U. Press, 1990).

[45] A.Huckelberry, T.Wurzbacher, \textit{Infinite Dimensional Kahler
Manifolds},

\ \ \ \ \ \ (Birkhauser, Boston, 1997).

\bigskip

\ \ \ \ \ \ 

\end{document}